\documentclass[sigconf,screen]{acmart}
\renewcommand\footnotetextcopyrightpermission[1]{} % removes footnote with conference information in first column
\usepackage{booktabs}
\usepackage{multirow}
\usepackage{makecell}
\usepackage{subfigure}
\usepackage[misc]{ifsym} 
\usepackage{balance}
\usepackage{listings}
\usepackage{color, xcolor}
\usepackage{algorithm2e}
\usepackage{diagbox}
\usepackage{tcolorbox}
\usepackage{stfloats}
\usepackage{enumitem}
\AtBeginDocument{%
  \providecommand\BibTeX{{%
    \normalfont B\kern-0.5em{\scshape i\kern-0.25em b}\kern-0.8em\TeX}}}

\setcopyright{acmcopyright}
\acmYear{2025}
\copyrightyear{2025}
\setcopyright{acmlicensed}

\begin{document}
\begin{sloppypar}
	
	\author{Lingzhe Zhang}
	\affiliation{%
		\institution{Peking University}
		\city{Beijing}
		\country{China}}
	\email{zhang.lingzhe@stu.pku.edu.cn}
	
	\author{Tong Jia$^{\ast}$}
	\affiliation{%
		\institution{Peking University}
		\city{Beijing}
		\country{China}}
	\email{jia.tong@pku.edu.cn}
	
	\author{Mingyu Wang}
	\affiliation{%
		\institution{Peking University}
		\city{Beijing}
		\country{China}}
	\email{mingyuwang25@stu.pku.edu.cn}
	
	\author{Weijie Hong}
	\affiliation{%
		\institution{Peking University}
		\city{Beijing}
		\country{China}}
	\email{hongwj@stu.pku.edu.cn}
	
	\author{Chiming Duan}
	\affiliation{%
		\institution{Peking University}
		\city{Beijing}
		\country{China}}
	\email{duanchiming@stu.pku.edu.cn}
	
	\author{Minghua He}
	\affiliation{%
		\institution{Peking University}
		\city{Beijing}
		\country{China}}
	\email{hemh2120@stu.pku.edu.cn}
	
	\author{Rongqian Wang}
	\affiliation{%
		\institution{Huawei Technologies Co., Ltd.}
		\city{Beijing}
		\country{China}}
	\email{wangrongqian2@huawei.com}
	
	\author{Xi Peng}
	\affiliation{%
		\institution{Huawei Technologies Co., Ltd.}
		\city{Hong Kong SAR}
		\country{China}}
	\email{pancy.pengxi@huawei.com}
	
	\author{Meiling Wang}
	\affiliation{%
		\institution{Huawei Technologies Co., Ltd.}
		\city{Shenzhen}
		\country{China}}
	\email{wangmeiling17@huawei.com}
	
	\author{Gong Zhang}
	\affiliation{%
		\institution{Huawei Technologies Co., Ltd.}
		\city{Shenzhen}
		\country{China}}
	\email{nicholas.zhang@huawei.com}
	
	\author{Renhai Chen}
	\affiliation{%
		\institution{Huawei Technologies Co., Ltd.}
		\city{Beijing}
		\country{China}}
	\email{chenrenhai@huawei.com}
	
	\author{Ying Li$^{\ast}$}
	\affiliation{%
		\institution{Peking University}
		\city{Beijing}
		\country{China}}
	\email{li.ying@pku.edu.cn}
	
	\renewcommand{\shortauthors}{Lingzhe Zhang et al.}

\title[Efficient Failure Management for Multi-Agent Systems with Reasoning Trace Representation]{EAGER: Efficient Failure Management for \\ Multi-Agent Systems with Reasoning Trace Representation}

\begin{abstract}
	Large Language Models (LLM)-based Multi-Agent Systems (MASs) have emerged as a new paradigm in software system design, increasingly demonstrating strong reasoning and collaboration capabilities. As these systems become more complex and autonomous, effective failure management is essential to ensure reliability and availability. However, existing approaches often rely on per-trace reasoning, which leads to low efficiency, and neglect historical failure patterns, limiting diagnostic accuracy. In this paper, we conduct a preliminary empirical study to demonstrate the necessity, potential, and challenges of leveraging historical failure patterns to enhance failure management in MASs. Building on this insight, we propose \textbf{EAGER}, an efficient failure management framework for multi-agent systems based on reasoning trace representation. EAGER employs unsupervised reasoning-scoped contrastive learning to encode both intra-agent reasoning and inter-agent coordination, enabling real-time step-wise failure detection, diagnosis, and reflexive mitigation guided by historical failure knowledge. Preliminary evaluations on three open-source MASs demonstrate the effectiveness of EAGER and highlight promising directions for future research in reliable multi-agent system operations.
\end{abstract}




\maketitle

\section{Introduction}

Multi-Agent Systems (MASs) powered by Large Language Models (LLMs) are rapidly emerging as a new paradigm in software systems design~\cite{xiao2025clslog, zhang2025survey2, duan2025logaction, pan2025omni, he2025llm, du2024multi, huself, li2025swe, zhang2025knowledge}. By decomposing complex tasks into coordinated agents, each capable of reasoning, planning, and interacting with tools or environments, these systems offer unprecedented flexibility and autonomy. Such MASs are increasingly applied across diverse domains, including software engineering~\cite{zhang2025survey, zhang2025agentfm, zhang2025scalalog, zhang2025thinkfl, zhang2025xraglog, he2025walk, he2025united, hong2025cslparser}, intelligent assistants~\cite{wang2024mobile, wen2025mica, xu2025comfyui, hu2025owl, zhang2025microremed}, and scientific workflows~\cite{ghafarollahi2025sciagents, gottweis2025towards, swanson2024virtual}, demonstrating their remarkable capability in handling complex, dynamic, and multi-step problems. 

However, the very capabilities that make LLM-powered MASs powerful also introduce unique operational challenges: their dynamic behavior and reasoning processes can lead to unpredictable failures, and traditional monitoring or debugging approaches are often inadequate to ensure efficient and accurate system operation. To address these challenges, the concept of Agent System Operations (AgentOps) has been proposed~\cite{wang2025survey}, offering a systematic framework for managing, diagnosing, and mitigating failures in multi-agent systems by leveraging agent-level reasoning traces alongside traditional monitoring data.

MAST~\cite{cemri2025multi} and TRAIL~\cite{deshpande2025trail} analyze failures in general-purpose MASs. Ruofan et al.~\cite{lu2025exploring} and Simiao et al.~\cite{liu2025empirical} focus on failures in code-based MASs. Xuyan et al.~\cite{ma2025diagnosing} examine failures in platform-orchestrated agentic systems. Meanwhile, Alfonso et al.~\cite{amayuelas2024multiagent}, TrustAgent~\cite{yu2025survey}, G-Safeguard~\cite{wang2025g}, and GUARDIAN~\cite{zhou2025guardian} primarily address anomaly detection, which determines whether a particular MAS reasoning process has failed or been compromised. In contrast, Who\&When~\cite{zhangagent}, AgenTracer~\cite{zhang2025agentracer}, AEGIS~\cite{kong2025aegis}, FAMAS~\cite{ge2025introducing}, A2P~\cite{west2025abduct}, and RAFFLES~\cite{zhu2025raffles} focus on failure diagnosis, identifying which agent (Who), at which step (When), experienced which error (What) that led to task failure. Although the effectiveness of these methods has been demonstrated in general-purpose MASs, they encounter several challenges when applied to a practical MAS.

\begin{itemize}[leftmargin=*]
	\item \textbf{Per-Trace Reasoning Leads to Low Efficiency.}  
	Reasoning traces generated by MASs are rich in semantic information. Current methods typically process each trace independently, relying on a large judge LLM to perform reasoning~\cite{zhang2025agentracer, west2025abduct, zhu2025raffles} for semantic anomaly detection as well as failure diagnosis. While effective, this approach is extremely time-consuming: not only does it analyze each trace individually, but the use of a large judge LLM further increases computational overhead. In a high-throughput MAS handling numerous requests, this combination results in significantly reduced operational efficiency.
	\item \textbf{Neglecting Historical Failure Patterns Limits Accuracy.}  
	While current methods leverage large judge LLMs to infer human-like solutions for novel failures based on the rich semantics of reasoning traces, these LLMs are inherently unstable. This instability means that the same failure may sometimes be analyzed correctly and sometimes incorrectly. Although methods such as RAFFLES~\cite{zhu2025raffles} introduce a set of specialized evaluators to assess the quality of the judge's reasoning, this merely adds layers of evaluation without addressing the underlying issue. Therefore, effectively leveraging historical failure patterns is crucial to ensure that previously encountered failures are recognized and handled reliably, thereby improving overall accuracy.
\end{itemize}

For complex general-purpose MASs, the aforementioned challenges are difficult to address due to the unpredictable and highly variable nature of failures~\cite{cemri2025multi, deshpande2025trail, zhang2026hypothesize, zhang2026agentic, liu2025ora, liu2024uac, liu2025aaad, huang2025uda}. In contrast, a specific practical MAS typically exhibits relatively fixed task types, stable agent invocation chains, and a consistent organizational structure. As a result, its failure types and patterns are comparatively limited. When AgentOps methods originally designed for general-purpose MASs are applied to such a specific system, historical reasoning traces can be leveraged to enable fast and accurate failure management, avoiding redundant analysis and mitigating the inherent instability of judge LLMs. However, constructing an effective representation model for reasoning traces involves several critical problems that must be addressed:

\underline{\textit{(P1) Representation Learning for Reasoning-Oriented Traces.}} To accelerate failure understanding, models must capture reasoning traces that fundamentally differ from conventional textual or operational representations. Each trace encodes both intra-agent reasoning dynamics and inter-agent coordination logic. Learning representations that preserve both scopes while maintaining discriminative power across diverse reasoning styles presents a significant challenge.

\underline{\textit{(P2) Lack of Labeled and Generalizable Failure Data.}} Improving diagnostic accuracy requires semantically rich supervision. Yet, labeled multi-agent failure data—particularly those with explicit root causes—are extremely scarce. Moreover, such data often exhibit strong domain bias, limiting the generalization of trained models. Consequently, developing a learning paradigm capable of leveraging unlabeled or weakly structured traces for reasoning-level alignment is essential.

Recognizing these challenge, we first conduct a preliminary empirical study to gain a deeper understanding of the underlying characteristics of failures in LLM-based MAS. Specifically, we aim to demonstrate that (i) failures within a fixed multi-agent system are often limited and recurring rather than arbitrary, suggesting strong potential for historical pattern reuse; and (ii) existing state-of-the-art text embedding models fail to effectively capture the reasoning structure and relational semantics inherent in agent reasoning traces, indicating the need for a dedicated representation learning paradigm.

Building on these insights, we propose \textbf{EAGER} — an \textbf{E}fficient failure management framework for multi-\textbf{AGE}nt systems with \textbf{R}easoning trace representation. EAGER models the reasoning trace generated during multi-agent collaboration, capturing both intra-agent reasoning dynamics and inter-agent orchestration patterns. A Representation Model, trained via Reasoning-Scoped Contrastive Learning, learns to encode these two reasoning scopes into a unified latent space, enabling efficient comparison and retrieval of similar reasoning experiences from historical traces. During system execution, EAGER performs Step-Wise Detection to identify potential failures at each reasoning stage. Once detected, it triggers a Reflexive Mitigation process, allowing agents to self-reflect, replan, or regenerate responses for rapid recovery. When a final output is confirmed as incorrect by the user, EAGER conducts Expert Inspect + Agent RCA to update both fine-grained (agent-level) and coarse-grained (system-level) failure knowledge, thereby continuously improving the system’s failure management ability.

We conduct preliminary experiments on three public open-source MAS to evaluate the effectiveness of EAGER in anomaly detection and failure diagnosis. Furthermore, its instantaneous detection and reflexive mitigation mechanisms can also enhance response accuracy in specific task.

\begin{figure*}[h]
	\begin{minipage}[b]{1.0\linewidth}
		\centering
		\includegraphics[width=\linewidth]{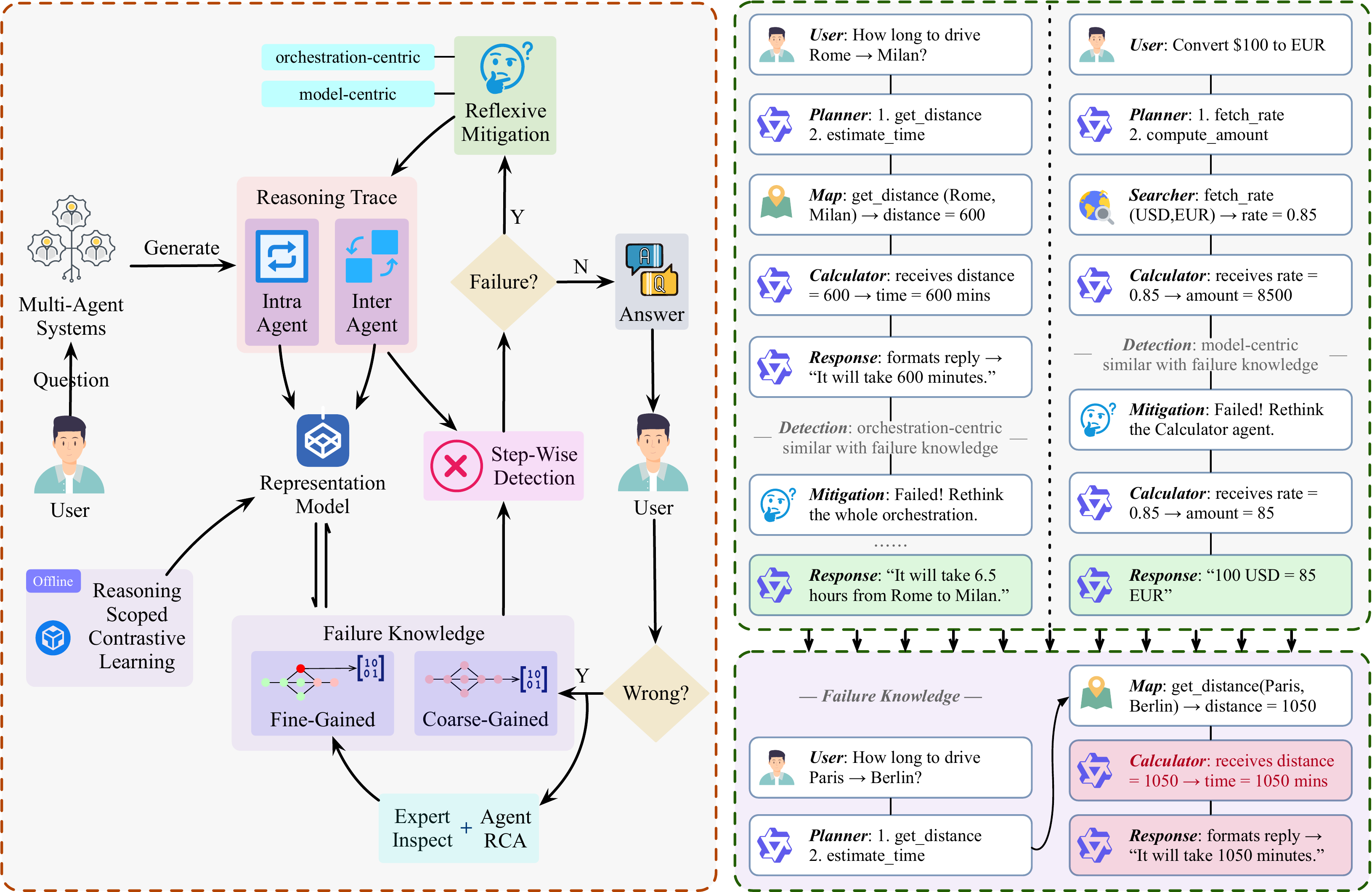}
		\vspace{-1.0em}
		\caption{Workflow of EAGER. The left panel shows the overall framework, while the right illustrates an example.}
		\label{fig: workflow}
		\vspace{-0.8em}
	\end{minipage}
\end{figure*}

\section{Preliminary Empircal Study}

\subsection{Failure Pattern Concentration in Multi-Agent Systems}

To investigate how failures distribute across different reasoning processes, we analyze three open-source MASs—AutoGen-Code~\cite{lu2025exploring}, RCLAgent~\cite{zhang2025adaptive}, and SWE-Agent~\cite{yang2024swe}. Using their official task datasets, we manually examine failed cases and categorize them into representative failure types.

\vspace{-0.5em}
\begin{table}[htb]
	\setlength{\tabcolsep}{2.5pt}
	\centering
	\caption{Failure Type Analysis across Three MAS}
	\vspace{-0.5em}
	\label{tab: failure-pattern-study}
	\begin{tabular}{c|ccc}
		\toprule
		\textbf{Failure Type} & \textbf{\textit{AutoGen-Code}} & \textbf{\textit{RCLAgent}} & \textbf{\textit{SWE-Agent}} \\
		\midrule
		Decomposition Error & 34.48\% & 0.00\% &  0.00\% \\
		Incorrect Code & 48.28\% & 0.00\% & 46.15\% \\
		Round Limitation & 17.24\% & 5.26\% &  0.00\% \\
		Critical Trace Miss & 0.00\% & 52.63\% &  0.00\% \\
		Metrics Query Error & 0.00\% & 42.11\% &  0.00\% \\
		Editing Error & 0.00\% & 0.00\% & 25.64\% \\
		Localization Error & 0.00\% & 0.00\% & 28.21\% \\
		\bottomrule
	\end{tabular}
\vspace{-0.8em}
\end{table}
\vspace{-0.5em}

As shown in Table~\ref{tab: failure-pattern-study}, the three MASs exhibit highly distinct failure distributions. AutoGen-Code mainly fails due to Incorrect Code and Decomposition Error, RCLAgent due to Critical Trace Miss and Metrics Query Error, while SWE-Agent often suffers from Editing and Localization Errors. These results reveal a clear concentration of failure patterns within specific reasoning or coordination scopes, suggesting that historical failure knowledge can be leveraged to enhance the efficiency and accuracy of MAS failure management.

\subsection{Evaluation of Existing Embeddings on Reasoning Trace Representation}

To assess whether current state-of-the-art (SOTA) embedding models can effectively represent reasoning traces, we conduct a retrieval-based experiment using two leading models: \textit{Qwen3-0.6B-Embedding}~\cite{yang2025qwen3} and \textit{BGE-M3-Embedding}~\cite{chen2024bge}. Specifically, we use reasoning traces generated by RCLAgent on nine distinct questions, each queried five times, yielding a total of 45 reasoning traces. Human annotators verify that the five traces corresponding to the same question are semantically similar.

\vspace{-0.5em}
\begin{table}[htb]
	\setlength{\tabcolsep}{3pt}
	\centering
	\caption{Similar Reasoning Trace Retrieval Experiment}
	\vspace{-0.5em}
	\label{tab: trace-retrieval}
	\begin{tabular}{c|ccc}
		\toprule
		\textbf{Model} & \textbf{\textit{Recall@10}} & \textbf{\textit{NDCG@10}} & \textbf{\textit{MRR@10}}\\
		\midrule
		Qwen3-0.6B-Embedding & 13.3\% & 8.7\% & 6.2\% \\
		BGE-M3-Embedding & 22.2\% & 14.5\% & 10.8\% \\
		\bottomrule
	\end{tabular}
\vspace{-0.8em}
\end{table}

As shown in Table~\ref{tab: trace-retrieval}, both models perform poorly in retrieving semantically similar reasoning traces. This suggests that current text embedding models—though highly effective in general NLP tasks—struggle to capture the complex hierarchical reasoning and coordination semantics inherent to multi-agent reasoning traces. These findings call for a representation method that models reasoning scope and agent interactions.

\section{Methodology}

Building on the findings of preliminary empirical study, we propose \textbf{EAGER}, an \textbf{E}fficient failure management framework for multi-\textbf{AGE}nt systems with \textbf{R}easoning trace representation. As shown in Figure~\ref{fig: workflow}, EAGER begins by capturing reasoning traces generated during multi-agent interactions, which contain both intra-agent reasoning processes and inter-agent orchestration patterns. A dedicated Representation Model, trained via \textit{Reasoning-Scoped Contrastive Learning}, learns to encode these traces into a unified latent space that aligns both intra-agent and inter-agent semantics.

During system operation, EAGER performs \textit{Step-Wise Detection}, which continuously evaluates each reasoning step against accumulated historical \textit{Failure Knowledge} to identify potential failures in real time. Upon detecting a failure, a \textit{Reflexive Mitigation} mechanism is triggered—allowing agents to self-reflect, replan, or regenerate responses—thereby effectively recovering from most failures. 

\begin{figure*}[h]
	\begin{minipage}[b]{1.0\linewidth}
		\centering
		\includegraphics[width=\linewidth]{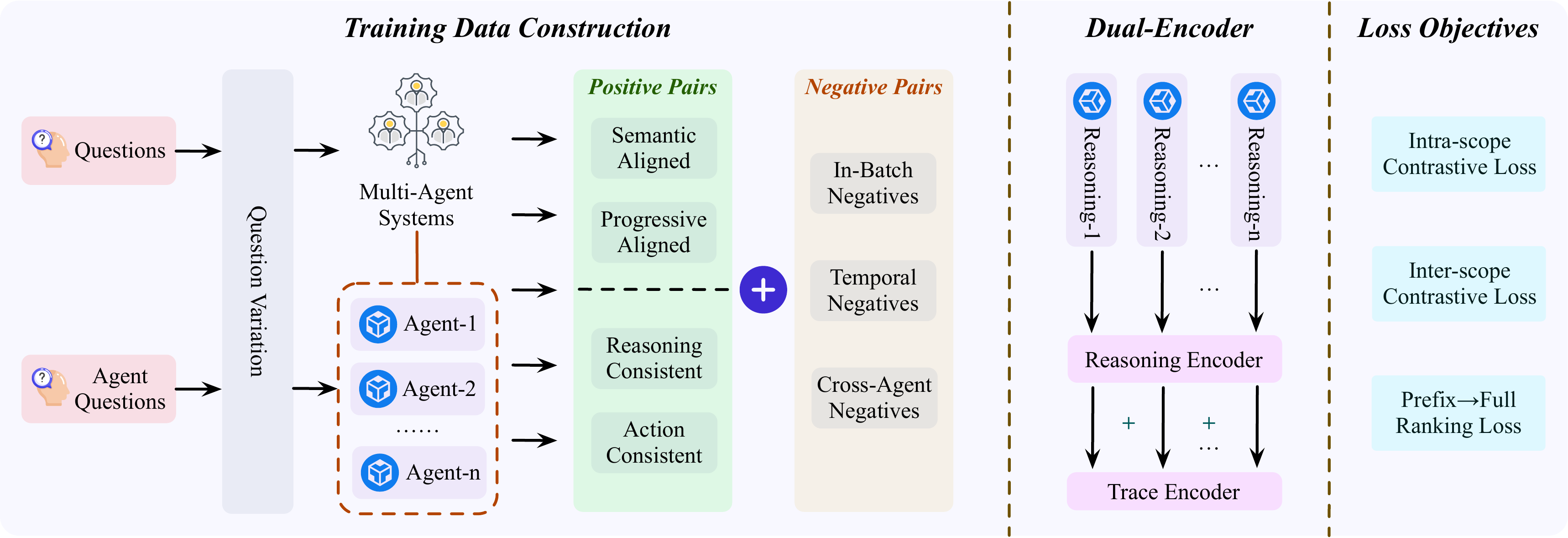}
		\vspace{-1.0em}
		\caption{Training Pipeline of Reasoning Scoped Contrsative Learning}
		\label{fig: training}
		\vspace{-0.8em}
	\end{minipage}
\end{figure*}

Finally, when the multi-agent system produces an output that the user deems incorrect, EAGER triggers an optional Expert Inspect + Agent RCA procedure. This process first employs SOTA AgentOps methods to analyze the corresponding reasoning trace, after which expert judgment is applied to validate and refine the findings. The resulting insights are used to enrich both fine-grained (agent-level) and coarse-grained (system-level) failure knowledge, which is then incorporated into the historical knowledge base to continually improve future detection and mitigation.

\subsection{Failure Management Components}

\paragraph{\textbf{Failure Knowledge.}} The failure knowledge is composed of both fine-grained knowledge and coarse-grained knowledge. \textit{Fine-grained knowledge} refers to the identification of specific reasoning errors in an individual agent's reasoning process. For example, it captures the specific reasoning steps or decisions within the agent's workflow where hallucinations, inconsistencies, or other errors occur. This knowledge is more granular and helps pinpoint the source of the problem at the agent level. \textit{Coarse-grained knowledge}, on the other hand, refers to an understanding of the overall failure at the trace level, where the entire reasoning trace is recognized as erroneous, but the exact agent or step responsible for the failure remains undetermined. This level of knowledge is crucial for quickly detecting failures without knowing the specific source.

\paragraph{\textbf{Step-Wise Detection.}} By leveraging failure knowledge for matching, anomaly detection becomes significantly faster. Therefore, EAGER performs step-wise detection, conducting a check after each agent completes its reasoning. Specifically, the reasoning embedding generated by the current agent is matched with the fine-grained knowledge. Once all agents have completed their reasoning, the entire reasoning trace is embedded and compared with the coarse-grained knowledge. If any similarity is detected, a fault is identified, triggering the reflexive mitigation process.

\paragraph{\textbf{Reflexive Mitigation.}} Most non-malicious failures in multi-agent systems can be self-corrected through reflective reasoning. Accordingly, EAGER incorporates a reflexive mitigation mechanism that operates at two complementary levels. When step-wise detection precisely identifies a specific agent’s failure, EAGER performs a model-centric reflection to refine the reasoning process of that agent. Conversely, when the entire reasoning trace is deemed faulty, EAGER triggers an orchestration-centric reflection, enabling the system to re-evaluate inter-agent coordination and restore consistent reasoning dynamics at the global level.

\subsection{Reasoning Scoped Contrsative Learning}

The training pipeline of reasoning-scoped contrastive learning is illustrated in Figure~\ref{fig: training}. To ensure a highly generalizable representation model, we intentionally avoid relying on failure-labeled data. Instead, we construct training samples through question variation, under the hypothesis that semantically similar questions tend to yield reasoning traces with analogous structures and logical progressions in most cases.

From this process, we extract both system-level reasoning traces (across all agents) and agent-level reasoning segments (from individual agents), which naturally provide positive and negative pairs required for contrastive learning. We then jointly train two hierarchical encoders: the \textit{Reasoning Encoder}, which captures intra-agent reasoning semantics, and the \textit{Trace Encoder}, which integrates multiple reasoning embeddings to encode inter-agent orchestration dependencies.

\vspace{-0.5em}
\begin{equation}
	\mathcal{L}{\text{total}} =
	\lambda_1 \mathcal{L}{\text{intra}} +
	\lambda_2 \mathcal{L}{\text{inter}} +
	\lambda_3 \mathcal{L}{\text{rank}},
	\label{eq: loss}
\end{equation}

The overall objective function $\mathcal{L}_{\text{total}}$ is composed of three complementary components, formulated as Equation~\ref{eq: loss}, where $\mathcal{L}_{\text{intra}}$ represents the \textbf{intra-scope contrastive loss}, enforcing proximity between reasoning embeddings of the same agent under similar question variations, $\mathcal{L}_{\text{inter}}$ denotes the \textbf{inter-scope contrastive loss}, aligning reasoning patterns across agents to preserve coordination semantics, and $\mathcal{L}_{\text{rank}}$ is the \textbf{prefix-to-full ranking loss}, which encourages consistency between partial and complete reasoning traces, enabling robust representation of reasoning trace.

\section{Preliminary Evaluation}

\begin{table}[htb]
	\setlength{\tabcolsep}{3pt}
	\centering
	\caption{Anomaly Detection \& Diagnosis Results}
	\vspace{-0.5em}
	\label{tab: anomaly-detection-diagnosis}
	\begin{tabular}{c|ccc}
		\toprule
		\textbf{Task} & \textbf{\textit{AutoGen-Code}} & \textbf{\textit{RCLAgent}} & \textbf{\textit{SWE-Agent}} \\
		\midrule
		Anomaly Detection & 73.57\% & 86.18\% & 79.95\% \\
		Failure Diagnosis & 63.23\% & 78.76\% & 69.51\% \\
		\midrule
		Detection Latency & 5.23s & 4.57s & 4.91s \\
		\bottomrule
	\end{tabular}
	\vspace{-0.8em}
\end{table}
\vspace{-0.5em}

To evaluate the feasibility and effectiveness of \textbf{EAGER}, we conduct experiments on the three aforementioned open-source MASs. The results for anomaly detection and failure diagnosis, evaluated by F1-Score, as well as the average detection latency, are presented in Table~\ref{tab: anomaly-detection-diagnosis}. EAGER demonstrates consistently strong performance across all tasks, highlighting its potential for efficient and accurate failure management in multi-agent systems. However, our current representation model is only lightly fine-tuned from the Qwen-0.6B-Embedding backbone, which limits its generalization capability. In future work, we plan to perform large-scale fine-tuning and further optimization to enhance both cross-domain robustness and reasoning adaptability.

\vspace{-0.5em}
\begin{table}[htb]
	\setlength{\tabcolsep}{3.5pt}
	\centering
	\caption{Task Performance Improvement with EAGER}
	\vspace{-0.5em}
	\label{tab: performance-improve}
	\begin{tabular}{c|ccccc}
		\toprule
		\textbf{Method} & \textbf{\textit{R1}} & \textbf{\textit{R3}} & \textbf{\textit{R5}} & \textbf{\textit{R10}} & \textbf{\textit{MRR}} \\
		\midrule
		RCLAgent & 28.47\% & 62.37\% & 64.41\% & 68.14\% & 46.13\% \\
		RCLAgent + EAGER & 30.19\% & 65.82\% & 68.56\% & 70.03\% & 48.65\% \\
		\bottomrule
	\end{tabular}
	\vspace{-0.8em}
\end{table}

Furthermore, since EAGER enables instantaneous detection and reflexive mitigation, it can also enhance the response accuracy of specific tasks. To verify this, we conducted additional experiments on RCLAgent, where task performance was evaluated using Recall@$k$ (R@$k$) and Mean Reciprocal Rank (MRR). As shown in Table~\ref{tab: performance-improve}, integrating EAGER with RCLAgent yields consistent and noticeable improvements across all evaluation metrics, demonstrating its effectiveness in enhancing real-time reasoning reliability.

\section{Conclusion}

This paper explores the necessity, potential, and challenges of leveraging historical failure patterns to enhance failure management in MASs. Building on these insights, we propose EAGER, an efficient failure management framework for MASs with reasoning-trace representation. Preliminary experiments demonstrate the feasibility and effectiveness of EAGER. In future work, we plan to further improve its generalization capability through large-scale fine-tuning.

\section*{Acknowledgment}

This work was supported by the Huawei–Peking University Joint Laboratory of Mathematics.

\bibliographystyle{ACM-Reference-Format}
\balance
\bibliography{sample-base}

\clearpage

\end{sloppypar}
\end{document}